\documentclass[twocolumn,showpacs,preprintnumbers,amsmath,amssymb]{revtex4}

\usepackage[dvips]{graphicx}
\usepackage{dcolumn}
\usepackage{bm}

\begin{document}

\title{Temperature evolution of infrared- and Raman-active phonons in graphite}

\author{P. Giura$^1$, N. Bonini$^2$, G. Creff$^3$, J.B. Brubach$^3$, P. Roy$^3$, and M. Lazzeri$^1$}
\affiliation{$^1$ IMPMC, Universit\'e Pierre et Marie Curie, CNRS, 4 place Jussieu, 75005 Paris, France}
\affiliation{$^2$ Dept. of Physics, King's College London, London WC2R 2LS, UK}
\affiliation{$^3$ Synchrotron SOLEIL, l'Orme des Merisiers, Saint-Aubin, BP 48, 91192 Gif-sur-Yvette, France}

\begin{abstract}

We perform a comparative experimental and theoretical study of the
temperature dependence up to 700 K of the frequency and linewidths
of the graphite $E_{1u}$ and $E_{2g}$ optical phonons ($\sim$ 1590
and 1580 cm$^{-1}$) by infra-red (IR) and Raman spectroscopy.
Despite their similar character, the temperature dependence of the
two modes is quite different, 
e.g., the frequency shift of the IR-active $E_{1u}$ mode
is almost twice as big as that of the Raman-active $E_{2g}$ mode.
{\it Ab initio} calculations of the
anharmonic properties are in remarkable agreement with measurements
and explain the observed behavior.

\end{abstract}


\maketitle Thermal properties of $sp^2$ carbon systems such as
graphene~\cite{balandin08,seol10} and carbon
nanotubes~\cite{yu05,pop06} have attracted significant attention,
both systems being excellent thermal conductors. In these materials,
thermal transport is dominated by lattice vibrations (phonons) and
transport can be described properly only once phonon
scattering mechanisms are taken into account. Phonon scattering also
plays a major role in non-equilibrium phenomena in which nonthermal
phonon populations of carbon materials are induced by optical
excitation~\cite{chatzakis11,scheuch11} or during electronic
transport~\cite{pop05,lazzeri06}. In general, our understanding of
thermal properties is heavily based on theoretical
modeling~\cite{bonini12,nika12,bonini07} and case studies allowing
to validate the models are of primary importance.

The frequency and the linewidth of a phonon, as well as their
temperature dependence, are measurable quantities which provide key
information about the interatomic interactions~\cite{menendez84}.
Indeed, even in a perfect crystal, the phonon linewidth is not zero
and the phonon frequencies depend on temperature because of
anharmonic terms in the interatomic potential (phonon-phonon
interaction) and, in the same instance, because of non-adiabatic
effects (phonon-electron interactions) when the gap is sufficiently
small. Raman spectroscopy certainly has a central role in the study
of vibrational properties of $sp^2$ carbon materials. However, this
technique probes only a very small portion of the vibrational
degrees of freedom and the use of complementary tools is essential
to gain further insight. Infrared (IR) spectroscopy is a natural
choice and it can also be used to study planar $sp^2$
systems~\cite{li12,manzardo12}.

The graphene in-plane anti-phase movement of the two unit-cell atoms
is a Raman-active zero momentum ({\bf q=0}) mode with symmetry
$E_{2g}$ (Raman $G$ peak). The $G$ frequency depends on the
temperature and the $G$ peak Raman measurement is currently used as
a probe of the local temperature in graphene
samples~\cite{balandin08,seol10}. In graphite, the $E_{2g}$ mode
splits in two modes with symmetry $E_{2g}$ and $E_{1u}$, depending
on the relative motion of the two graphene planes of the graphite
unit cell~\cite{tuinstra70}. The $E_{2g}$ one  (in-phase vibration
of the two planes) is Raman active and is associated with the well
known $G$ peak at $\sim$1582 cm$^{-1}$. The $E_{1u}$ one (anti-phase
vibration of the two planes) is IR active and has a slightly higher
frequency ($\sim$1590 cm$^{-1}$). The anomalously large IR intensity
of this mode has been the topic of a recent work~\cite{manzardo12}.

Given the similarities between the $E_{2g}$ and $E_{1u}$ vibrations,
their properties are expected to be similar. On the contrary, their
linewidths are quite different: At room conditions, the full-width
at half maximum (FWHM) of the Raman mode is $\sim$13 cm$^{-1}$,
while the FWHM of the IR one is $\sim$4 cm$^{-1}$. Based on
calculations~\cite{bonini07} these width differences have been
related to the different electron-phonon interaction. Moreover,
according to Ref. ~\cite{bonini07}, the Raman mode linewidth
decreases by increasing the temperature (this behavior is quite
unusual), while the IR linewidth is predicted to increase. Above
room temperature, measurements of the temperature dependence of
the linewidth and frequency shift of the IR mode are needed
for a comparison.

In this Rapid Communication, we present the measurement of the line-width
and the line-shift of the infra-red active mode at $\sim$ 1590
cm$^{-1}$ in highly oriented pyrolitic graphite in the temperature
range 293-523 K. For a comparison, we also show Raman spectroscopy
measurements of the $G$ peak in a similar temperature range. The
results are interpreted by means of {\it ab-initio} calculations of the
anharmonic properties.

\begin{figure} [ht]
\includegraphics[width=6.5cm]{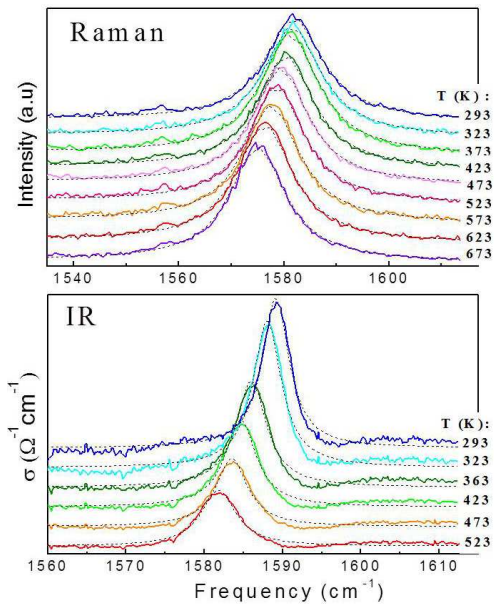}
\caption{(Color online) Graphite measured Raman spectra and IR optical
conductivity spectra. Spectra taken at different temperatures ($T$) are
vertically shifted for clarity.
Each peak is fitted with a single Lorentzian (thin dotted lines).}
\label{fig1}
\end{figure}

Raman spectra were measured using a Horiba Jobin Yvon T64000
spectrometer equipped with an Ar laser (514.5 nm). The laser was
focused on the sample via a Mitutoyo 20$\times$ long distance objective.
The incoming power was selected in order to prevent sample damage
and ensure a rather high signal to noise ratio. The instrumental
resolution obtained by measuring the emission spectrum of neon in
the $E_{2g}$ phonon frequency region has been found to be 1.6
cm$^{-1}$ (FWHM of a Gaussian profile). Spectra were recorded by
spanning the region (1200-1700 cm$^{-1}$). The sample was mounted
inside an in-house made oven composed of a cartridge heater tightly
inserted in a cylindrical copper piece with a little squared open
cavity for the incoming light. The heater was connected to a power
supply equipped with a thermometer allowing to control the
temperature of the cartridge heater. The precision was $\pm$0.1ºC.

IR spectra were measured at the AILES infrared
beamline~\cite{braubach} of synchrotron SOLEIL (Saint Aubain,
France) using a Bruker IFS125 spectrometer working under vacuum. The
temperature was set by using a in house made oven similar to the
one used for the
Raman experiments. The sample was aligned in reflection geometry
allowing to acquire reflectivity measurements of the $ab$-plane
transverse excitations at increasing temperatures.  The reference
was measured at room temperature using a gold mirror mounted in such
a way as to replace the sample inside the oven. The entire working
set-up used was as follows: synchrotron light as the source, KBr as
the beam splitter
and mercury cadmium telluride (MCT) as the detector. With this
configuration we acquired spectra in the frequency region between
700 and 3500 cm$^{-1}$ with a resolution of 0.1 cm$^{-1}$. Reflectivity
data were treated by Kramers-Kronig transformation to extract the
real part of the optical conductivity $\sigma(\omega)$. The Raman
spectra and the conductivity IR spectra, corrected by the respective
background, were fitted with a single Lorentzian profile
(Fig.~\ref{fig1}) to obtain the width and position of the peaks. For a
discussion on the asymmetric shape of the IR peak see
~\cite{manzardo12}.

\begin{figure} [h]
\includegraphics[width=7.5cm]{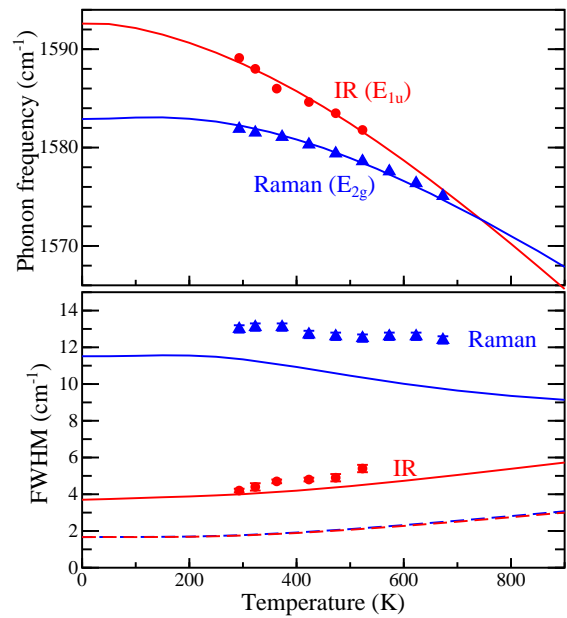}
\caption{(Color online) Upper: Graphite phonon frequency as a
function of temperature. Symbols are the present measurements from
Raman (triangles) and IR (dots). Solid lines are calculations. Lower:
Linewidths (full width at half maximum) corresponding to the
upper panel measurements. Dashed lines are the contribution to the
linewidth due to anharmonic phonon-phonon scattering. Continuous
lines are the total intrinsic linewidth (phonon-phonon +
electron-phonon, see the text). The calculated Raman shift and the
calculated Raman and IR linewidths are taken from
Ref.~\cite{bonini07}.} \label{fig2}
\end{figure}

In a defect-free perfect crystal, phonons have a finite lifetime
because they can decay by anharmonic phonon-phonon ($ph-ph$)
scattering or by electron-phonon ($e-ph$) scattering, this last
process being possible in graphite as the electronic gap is zero.
The intrinsic linewidth of a phonon can thus be written as a sum of
two terms $\gamma(T)=\gamma^{ph-ph}(T)+\gamma^{e-ph}(T)$ (see
e.g.~\cite{bonini07} and references therein), where $T$ is the
temperature. The linewidth in a real sample can be larger because of
the presence of other scattering mechanisms. Concerning the
temperature dependence of the phonon pulsation $\omega_{ph}$, at the
lowest order one can distinguish the following contributions
~\cite{menendez84}:
\begin{equation}
\omega_{ph}=\omega_0 + \Delta\omega^{le}(T)+\Delta\omega^{3p}(T)+\Delta\omega^{4p}(T)+\mathcal{O}(\hbar^2),
\label{eq1}
\end{equation}
where $\omega_0$ is the harmonic pulsation at the
equilibrium lattice parameters,
$\Delta\omega^{le}(T)$
accounts for the variation of the harmonic pulsation by varying the
lattice parameters (as a consequence of the thermal expansion),
and $\Delta\omega^{3p}(T)$ and
$\Delta\omega^{4p}(T)$ can be interpreted as due to anharmonic scattering involving,
respectively, three or four phonons (see e.g.~\cite{menendez84,lazzeri03} and references therein).
The three  shifts are proportional to $\hbar$ and are, thus, expected
to be of the same order of magnitude. Other terms are $\mathcal{O}(\hbar^2)$.

In the present study, we compare the measured linewidths with those
calculated in~\cite{bonini07}. Reference~\cite{bonini07} reports the
calculated anharmonic lineshift for the Raman mode only. Here, we
determine the anharmonic lineshift of the IR active mode using the
same approach as ~\cite{bonini07}. Calculations are performed within
density functional theory (plane-waves and pseudopotential approach)
with the QUANTUM ESPRESSO package~\cite{giannozzi}. Phonon
dispersions are obtained within the linear response approach
of~\cite{DFPT}. Anharmonic phonon-phonon scattering coefficients are
determined thanks to the 2n+1 theorem as implemented
in~\cite{lazzeri02}. All the computational details have been
described in ~\cite{bonini07}, and, in particular, we use the local density
approximation and equilibrium lattice spacing parameters
($a=2.43$\AA, $c/a=2.725$).

\begin{figure} [ht!]
\includegraphics[width=7.5cm]{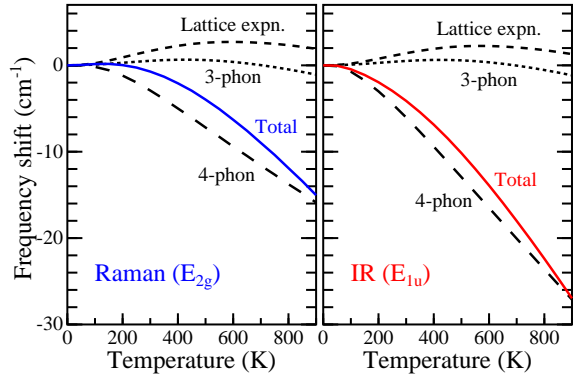}
\caption{(Color online) Calculated lineshift of the $E_{2g}$ (Raman active)
and $E_{1u}$ (IR active) ${\bm \Gamma}$ phonons of graphite.
The total lineshift (continuous line) is decomposed into the three-phonon,
four-phonon and lattice expansion contributions.
All the contributions are vertically shifted so as to be zero for $T$=0K.
The $T$=0K values are given in the text.
Calculations for the Raman mode are taken from Ref.~\cite{bonini07}.
}
\label{fig3}
\end{figure}

Figure~\ref{fig2} compares the measured and calculated parameters for
the Raman ($E_{2g}$) and IR active ($E_{1u}$) modes. Measured
linewidths are in good agreement with the calculations
of~\cite{bonini07}: The IR linewidth increases when increasing the
temperature while the Raman decreases. These trends are directly
visible on the raw data of Fig.~\ref{fig1}. This behavior was
interpreted in~\cite{bonini07} considering the following: (i) The
phonon-phonon contribution to the linewidth, $\gamma^{ph-ph}$, is
very similar for the two modes and increases with the temperature
(Fig.~\ref{fig1}); and (ii) the $\gamma^{e-ph}$ contribution is
significantly more important for the $E_{2g}$ mode  and
provides the overall temperature decrease of the linewidth.


Figure~\ref{fig2} also shows the line-shift. The theoretical curves
are vertically shifted to best fit the experimental data, obtaining,
for $T$=0K, 1592.6 and 1582.9 cm$^{-1}$ for $E_{1u}$ and
$E_{2g}$, respectively. These can be considered as the extrapolated
$T$=0K experimental frequencies and their difference is 9.7 cm$^{-1}$.
At the equilibrium lattice spacing,
the $E_{1u}$ ($E_{2g}$) frequency is 1614.2 (1604.9) cm$^{-1}$.
To compare
with the $T$=0K experimental value, one has to add the calculated
anharmonic shifts at $T$=0K, which are 
$\Delta\omega^{le}(0)=-17.7$ (-17.7) cm$^{-1}$,
$\Delta\omega^{3p}=-14.2$ (-14.1) cm$^{-1}$
and $\Delta\omega^{3p}=-1.3$ (+1.3) cm$^{-1}$ for
the $E_{1u}$ ($E_{2g}$) mode. The resulting frequency difference is
6.6~cm$^{-1}$, which underestimates by 3 cm$^{-1}$ the experimental
9.7 cm$^{-1}$ value. By considering frequencies at the 
generalized gradient approximation (GGA)
level,~\cite{mounet05} the agreement slightly worsens.

By increasing the temperature from 293 to 523 K, the measured IR
frequency shifts by $-7.3$~cm$^{-1}$, which is more than twice the
corresponding Raman shift, $-3.3$~cm$^{-1}$, in agreement with
calculations (Fig.~\ref{fig2}). To understand this behavior, we
decompose the lineshift of the two modes into the three $T$
dependent components defined in Eq.~\ref{eq1} (see Fig.~\ref{fig3}).
The different behavior of the two $E_{2g}$ and $E_{1u}$ modes is
almost entirely determined by the corresponding 4-phonon scattering
terms (Fig.~\ref{fig3}).

\begin{figure*}[ht!]
\center{
\includegraphics[width=3.8cm]{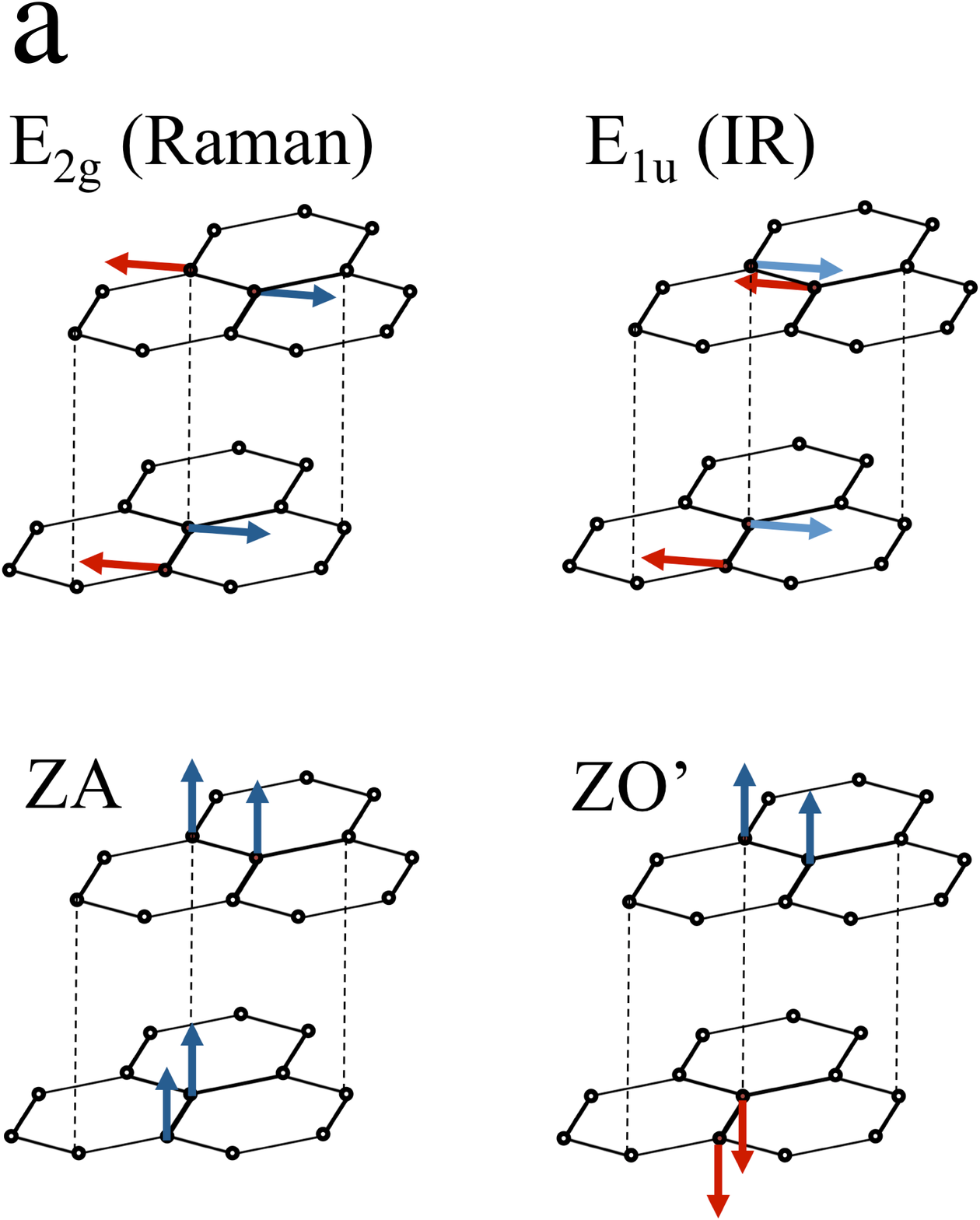}
\includegraphics[width=13.2cm]{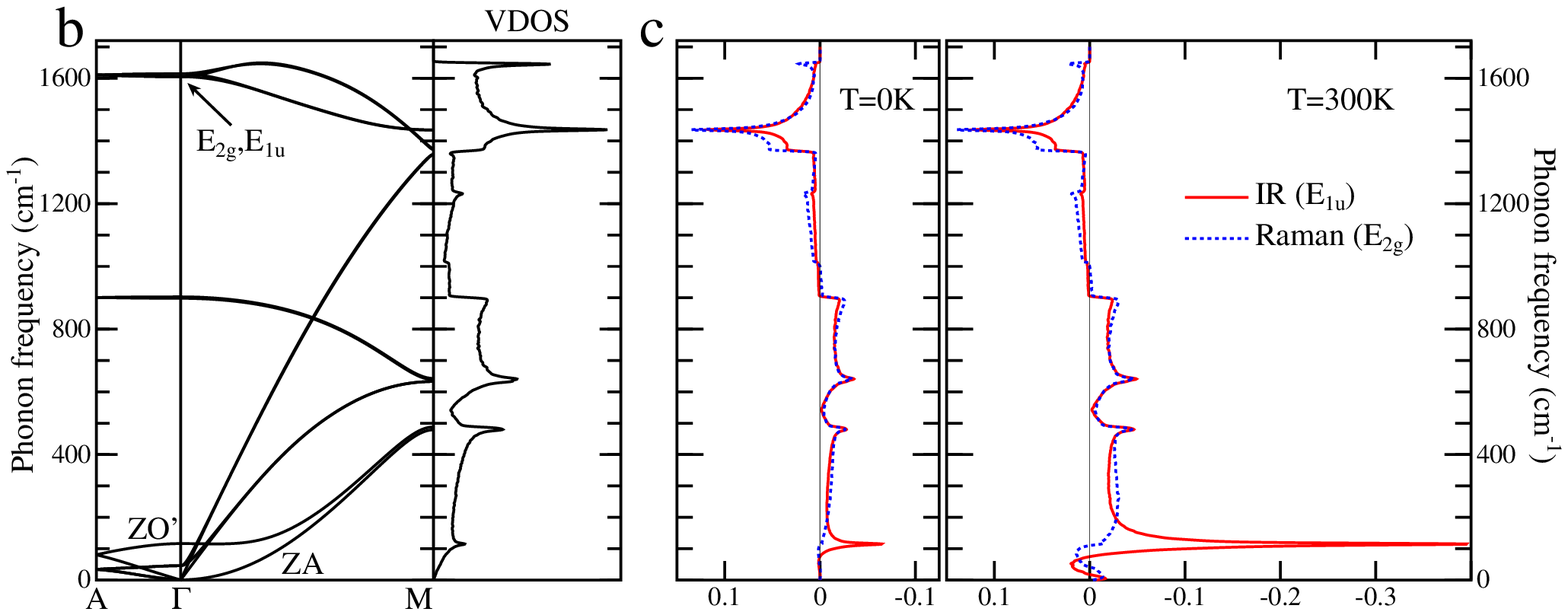}}
\caption{(Color online) (a): Scheme of various $ {\bm \Gamma}$ ({\bf q}={\bf 0}) phonons
of graphite. (b): Graphite phonon dispersion and
  related vibrational density of states (VDOS). (c): Spectral
  decomposition of the four-phonon component of the frequency shift
[as defined in Eq.~(\ref{eq2})] of
  the $E_{2g}$ (Raman) and $E_{1u}$ (IR) phonons at temperatures 0 and
  300 K.}
\label{fig4}
\end{figure*}

Why are the four-phonon shifts of the two modes so different?
To answer this question, we need to introduce some concepts.
Let us consider the interatomic potential energy
${\cal E}^{tot}(\{u_{s\alpha}({\bf R}_l)\})$,
where $u_{s\alpha}({\bf R}_l)$ is the
displacement from the equilibrium position of the $s$-th atom in the crystal
cell identified by the lattice vector ${\bf R}_l$ along the $\alpha$
Cartesian coordinate.
For a phonon with wavevector {\bf q},
branch index $i$, and energy $\hbar\omega_{{\bf q}i}$,
we define the adimensional phonon displacement as
$
u_{{\bf q}i}=1/N\sum_{l,s,\alpha}
\sqrt{2M_{s}\omega_{{\bf q}i}/\hbar}
v_{s\alpha}({\bf q}i) u_{s\alpha}({\bf R}_l) e^{-i{\bf q}\cdot{\bf R}_l}
$,
$v_{s\alpha}({\bf q}i)$ being the orthogonal phonon eigenmodes normalized
on the unit cell, $M_s$ the atomic mass, and $N$ the number of {\bf q}-points
describing the system (or unit cells).

For a phonon {\bf q=0}, and branch index $i$, we define the four-phonon spectral function as:
\begin{widetext}
\begin{equation}
S^{4p}_{{\bf 0}i}(\omega)=\frac{1}{2N\hbar}
\sum_{{\bf q}j}
\frac{\partial^4{\cal E}^{cell}}
{\partial u_{{\bf 0}i}\partial u_{{\bf 0}i}\partial u_{-{\bf  q}j}\partial u_{{\bf q}j}}
(2n_{{\bf q}j}+1)\delta(\omega-\omega_{{\bf q}j}),
\label{eq2}
\end{equation}
\end{widetext}
where $\sum_{\bm q}$ is a sum on the Brillouin zone,
${\cal E}^{cell}$ is the unit cell energy,
$n_{{\bf q}j}$ is the Bose-statistics occupation of the
phonon $({\bf q}j)$, and $\delta$ is the Dirac distribution.
Equation (\ref{eq2}) depends on the temperature only through the terms
$n_{{\bf q}j}$.
The associated four-phonon shift~\cite{menendez84,lazzeri03} can be written as
\[
\Delta\omega^{4p}_{{\bf 0}i}=\int_0^\infty~d\omega~S^{4p}_{{\bf 0}i}(\omega).
\]
The fourth order derivatives in Equation (\ref{eq2}) can be interpreted as
scattering coefficients among two phonons {\bf 0}$i$ and the two
phonons -{\bf q}$j$,{\bf q}$j$.
The spectral function separates the contributions from phonons
with different energies.

The spectral functions of the two $E_{2g}$ and $E_{1u}$ modes are
very similar [Fig.~\ref{fig4}(c)], the major differences being near
115 cm$^{-1}$, where the $E_{1u}$ spectral function shows a peak
absent in the $E_{2g}$ one. This peak strongly depends on the
temperature (in the temperature range considered) and accounts for
the difference in the two $E_{2g}$ and $E_{1u}$ shifts. This peak is
associated with the maximum of the graphite phonon density of states
(VDOS in Fig.~\ref{fig4}) at 115 cm$^{-1}$, and this maximum is due to
the optical out of plane $ZO'$ phonon branch. We remind that, in
graphene, the vibration of the atoms perpendicular to the
plane, is associated with the $ZA$ acoustic branch.
In graphite, this branch, splits into a branch in which the
two graphene planes vibrate in-phase, $ZA$, and a branch in which the
two planes vibrate anti-phase, $ZO'$, [Fig.~\ref{fig4}(a)-(b)]. The $ZA$
graphite branch is acoustic, while the $ZO'$ is optical with a 114
cm$^{-1}$ frequency at {\bf q=0}. The four-phonon anharmonic coupling
[the fourth order derivative in Eq. (\ref{eq2})] between the $E_{1u}$
mode and the $ZO'$ branch is much stronger than that between the
$E_{2g}$ and the $ZO'$. The presence (absence) of this coupling
explains the presence (absence) of the 115 cm$^{-1}$ peak in the
spectral decomposition of Fig.~\ref{fig4}c for the
$E_{1u}$($E_{2g}$) mode.

We remark that
graphene bilayer also presents an IR active $E_{1u}$ optical mode
and a $ZO'$ mode with a frequency of $\sim$ 80 cm$^{-1}$~\cite{park08},
smaller than in graphite.
A smaller frequency is associated with a stronger temperature dependence
of the $ZO'$ phonon occupation (with respect to the graphite $ZO'$).
This fact, suggests that the IR active $E_{1u}$ mode of the bilayer should present
a temperature dependence of the shift stronger than in graphite.

To conclude, we measured the linewidth and frequency shift of the
${\bm \Gamma}$ optical phonons $E_{1u}$ (IR active) and $E_{2g}$
(Raman active) of graphite as a function of temperature. Despite
the similarities between the two $E_{1u}$ and $E_{2g}$ vibrations,
their temperature dependence is quite different. The $E_{1u}$
linewidth is almost three times smaller than the $E_{2g}$ one (at
room temperature) and, while the $E_{1u}$ linewidth increases by
increasing the temperature, the $E_{2g}$ one slightly decreases.
Both behaviors are in good agreement with the {\it ab initio} calculations
of~\cite{bonini07}. The shifts of the two modes are also very
different: By increasing the temperature from 293 to 523 K, the
$E_{1u}$ mode shifts by more than twice the corresponding $E_{2g}$
value, in agreement with the present calculations. This difference
is explained by the presence (absence) of the  four-phonon anharmonic
coupling between the $E_{1u}$ ($E_{2g}$) phonon and the $ZO'$ phonon
branch at $\sim$ 115 cm$^{-1}$. These findings confirm the accuracy
and predictive power of density functional theory calculations in
determining anharmonic phonon-phonon scattering in $sp^2$ carbon
systems, a key step in developing realistic models for thermal
transport ~\cite{bonini12}.

Part of the calculations were done at IDRIS, project 096128.

\end{document}